\documentclass[11pt]{article}
\usepackage{mathrsfs,amssymb,axodraw,graphics}

\title{\bf Perturbative vs Schwinger-propagator method
for the calculation of amplitudes in a magnetic field}

\author{\bf Jos\'e F. Nieves\\
Laboratory of Theoretical Physics\\ 
Department of Physics, P.O. Box 23343\\
University of Puerto Rico, R\'{\i}o Piedras,
Puerto Rico 00931-3343
\and
\bf Palash B. Pal\\
Saha Institute of Nuclear Physics\\ 
1/AF Bidhan-Nagar, Calcutta 700064, India
}
\date{}

\evensidemargin=\oddsidemargin
\makeatletter
\@addtoreset{equation}{section}

\makeatother

\renewcommand{\slash}[1]{#1\llap/}
\def\ket#1{\left|#1\right>}
\def\bar{\overline}
\def\lag{\mathscr L}
\def\Tr{\mathop{\rm Tr}}
\def\ampl{\mathbb M}
\newcommand{\Eq}[1]{Eq.~(\ref{#1})}
\newcommand{\Eqs}[2]{Eqs.\ (\ref{#1}) and (\ref{#2})}
\newcommand{\Eqss}[3]{Eqs.\ (\ref{#1}), (\ref{#2}) and (\ref{#3})}
\newcommand{\Ref}[1]{Ref.~\cite{#1}}
\newcommand{\fig}[1]{Fig.~\ref{#1}}
\def\grav{{\cal G}}
\def\Graviton(#1,#2)(#3,#4)#5#6{%
\Photon(#1,#2)(#3,#4)#5#6
\Photon(#1,#2)(#3,#4){-#5}#6}

\begin{document}

\maketitle

\begin{abstract}
We consider the calculation of amplitudes for processes that take
place in a constant background magnetic field, first using the
standard method for the calculation of an amplitude in an external
field, and second utilizing the Schwinger propagator for charged
particles in a magnetic field.  We show that there are processes for
which the Schwinger propagator method does not yield the total
amplitude.  We explain why the two methods yield equivalent results in
some cases and indicate when we can expect the equivalence to hold. We
show these results in fairly general terms and illustrate them with
specific examples as well.
\end{abstract}

\section{Introduction}
In this paper we are concerned with the calculation of amplitudes for
processes that take place in a constant background magnetic field $B$.
There are many papers in the literature in which this kind of process
are considered \cite{Esposito:1995db, Elmfors:1996gy,
Ioannisian:1996pn, Erdas:1998uu, Ganguly:1999ts, Bhattacharya:2001nm,
Bhattacharya:2002aj, Tinsley:2001kb, Nieves:2003kw,
Bhattacharya:2003hq, Nieves:2004qp}, a significant fraction of which
have to do with neutrino processes that may take place in a variety of
astrophysical environments. It is useful to keep those particular
situations in mind, but for our purposes it is convenient to setup a
more general framework.

Thus, let us suppose that we want to calculate the amplitude
for the transition
\begin{eqnarray}
\ket i \stackrel B\longrightarrow \ket f \,,
\label{i->f}
\end{eqnarray}
where $\ket i$ and $\ket f$ denote two states, and the
letter $B$ above the arrow indicates that the
transition takes place in the presence of the external magnetic field.
For reasons that will become clear below, we restrict the initial and
final states to contain no charged particles, only neutral ones,
and we consider the calculation of the amplitude up to terms linear in $B$.

There are at least two ways to proceed with such a calculation.
One way is to start by computing in standard Feynman diagram
perturbation theory the Green's function for
\begin{eqnarray}
\ket {i + \gamma} \to \ket f \,,
\label{igf}
\end{eqnarray}
in the absence of $B$, and in particular with the photon off-shell,
which corresponds to the matrix element $\langle f|j_\mu|i\rangle$ of
the electromagnetic current.  Since the external particles are assumed
to be neutral, then in the context of the perturbative calculation the
off-shell photon is attached only to the internal lines of any diagram
that contributes to the Green's function.  The amplitude for the
transition in \Eq{i->f} is then obtained at the end by inserting the
appropriate photon field that corresponds to the background magnetic
field, together with the wavefunctions of the particles in the initial
and final states involved in the transition.  This procedure yields
the amplitude to first order in $B$.  This was in fact the approach
employed in the original calculation of the neutrino index of
refraction in the presence of a magnetic field \cite{D'Olivo:1989cr}.
We will refer to this method as the \emph{Perturbative Method}, or P
method for short.

An alternative approach is to calculate the Green's function
for the transition
\begin{eqnarray}
\ket i \rightarrow \ket f \,,
\end{eqnarray}
but employing the Schwinger \cite{Schwinger:1951nm}
propagator in place of the Feynman propagator for all the
charged internal lines that appear in any diagram that contributes
to the amplitude. Then, if the result is expanded in powers of $B$,
the conventional expectation is that the linear term in $B$ should
coincide with the result of the P method as specified above.
We will refer to the procedure that we have just described
as the \emph{Linear Schwinger Method}, or S method for short.

Indeed, the familiar calculations already mentioned involving
neutrino processes, confirm this expectation that both methods
yield the same result. The question we address here is
whether this is a special feature of the processes that have been
considered, and whether we can expect the result to hold 
for any other process of the type we are considering.

The purpose of this paper is to show that
there are processes for which this equivalence does not hold. 
To be specific, in those cases, the diagrams that contribute to the
amplitude in the P method can be classified in two topologically distinct
groups that we call type-1 and type-2 diagrams, which are distinguished
according to whether the electromagnetic vertex has only internal
lines attached to it, or whether it has some external lines attached as well.
As we show, the S method yields a result that is equivalent to the
result of the type-1 diagrams.
For processes for which the type-2 diagrams do not exist,
both the P and S methods yield the same result, which is the case
of the neutrino processes that we have mentioned. But in the
more general case in which the type-2 diagrams exist, the S
method does not yield the complete amplitude. The
total amplitude is obtained by taking the result of the type-1 diagrams,
which can be calculated by either method, and then adding
the result of the type-2 diagrams using the P method.

The paper is organized as follows.  In Sec.~\ref{s:linsch}, we derive
the linearized forms of the Schwinger propagators for fermions,
scalars and vector particles in order to set up the stage for
calculations in the S method.  In Sec.~\ref{s:ptb}, we outline the P
approach, introducing the classifications of all relevant diagrams
into two types, which we have called type-1 and type-2 above.  In
Sec.~\ref{s:equiv}, we show the example of a process in which the S
and the P methods yield identical results because only type-1 diagrams
are present.  In Sec.~\ref{s:noneq}, we discuss processes where type-2
diagrams are also present, so that the S method does not give the
total amplitude.  Finally, in Sec.~\ref{s:conclu}, we present our
conclusions.

\section{Linear Schwinger approach}\label{s:linsch}
The Schwinger formula \cite{Schwinger:1951nm} gives the fermion
propagator to all orders in the $B$ field, and the analogous formulas
for the scalar and vector propagators are also known
\cite{Erdas:gy}. Since we will be looking at the amplitudes calculated
to first order in $B$, a simpler formula, which is correct to that
order, is sufficient for our purpose. Below we give a short derivation
of the linear formulas for the propagators, in a manner that will be
useful in what follows.

\subsection{Fermion propagator in a magnetic field}
We follow the derivation given in \Ref{Nieves:2004qp}
for the fermion propagator,
and subsequently extend it to obtain the analogous result
for the scalar and vector propagator.
The propagator of a Dirac particle of mass $m$ and charge $eQ$
in an external electromagnetic field is
\begin{eqnarray}
\Big[ i \gamma^\mu {\cal D}_\mu - m \Big] S(x,x') 
= \delta^4 (x - x') \,,
\label{coordeq}
\end{eqnarray}
where the the electromagnetic gauge covariant (EGC) derivative ${\cal
D}_\mu$ is defined by
\begin{eqnarray}
{\cal D}_\mu = {\partial \over \partial x^\mu} + ieQ A_\mu(x)\,.
\label{Dmu}
\end{eqnarray}
The Schwinger solution is of the form 
\begin{eqnarray}
S(x,x') = \Phi(x,x')\int\,\frac{d^4 p}{(2\pi)^4}
e^{-ip\cdot(x - x')} S_F(p) \,,
\label{ansatz}
\end{eqnarray}
where the overall phase $\Phi(x,x')$ is chosen such that
\begin{eqnarray}
\label{Phieq}
i{\cal D}_\mu \Phi(x,x^\prime) = \frac{eQ}{2}F_{\mu\nu}(x - x^\prime)^\nu
\Phi(x,x^\prime) \,,
\end{eqnarray}
and it depends on the gauge choice for the background field.
For a constant field $F_{\mu\nu}$ corresponding to the
background magnetic field, we can choose the gauge for the
vector potential such that
\begin{eqnarray}
A_\mu(x) = -\frac{1}{2}F_{\mu\nu}x^\nu \,,
\label{Amu}
\end{eqnarray}
With this choice,
\begin{eqnarray}
\Phi(x,x') = \exp \Big( \frac{i}{2}eQ x^\mu F_{\mu\nu}x'^\nu \Big) \,,
\label{defphi}
\end{eqnarray}
and in a different gauge, the expression for $\Phi(x,x')$ will be
different \cite{Bhattacharya:2004nj, Bhattacharya:2005zu}.  We will
always employ the gauge dictated by \Eq{Amu}.

By virtue of \Eq{Phieq}, $\Phi$ has the property that, for an arbitrary
function $f$,
\begin{eqnarray}
{\cal D}_\mu (\Phi f) = \Phi\left[\partial_\mu -\frac{ieQ}{2}F_{\mu\nu}
(x - x^\prime)^\nu\right]f \,.
\end{eqnarray}
Moreover, if $f$ is a translationally invariant function of two
co-ordinates, $f(x,x')$, so that its Fourier transform is given by
\begin{eqnarray}
f(x,x') = \int {d^4p \over (2\pi)^4} e^{-ip\cdot(x - x')} \widetilde
f(p) \,,
\end{eqnarray}
then it follows that
\begin{eqnarray}
{\cal D}_\mu (\Phi f) = -i \Phi(x,x') \int {d^4p \over (2\pi)^4}
e^{-ip\cdot(x - x')} \widetilde{\cal D}_\mu \widetilde f(p) \,, 
\label{Dphif}
\end{eqnarray}
where
\begin{eqnarray}
\widetilde{\cal D}_\mu = p_\mu - \frac{ieQ}{2}F_{\mu\nu}
\frac{\partial}{\partial p_\nu} \,.
\label{Dp}
\end{eqnarray}

Substituting \Eq{ansatz} into \Eq{coordeq} and using \Eq{Dphif}, 
the equation for the momentum space propagator $S_F(p)$ is
\begin{eqnarray}
\Phi(x,x')\int\,\frac{d^4 p}{(2\pi)^4}
e^{-ip\cdot(x - x')} \left[\slash{p} - \frac{ieQ}{2}F^{\mu\nu}
\gamma_\mu\frac{\partial}{\partial p^\nu} - m\right]S_F(p) = {\delta^4
  (x - x')}  \,.
\end{eqnarray}
Using $\Phi(x,x)=1$ this equation can be solved by setting
\begin{eqnarray}
\left[\slash{p} - m - \frac{ieQ}{2}F^{\mu\nu}
\gamma_\mu\frac{\partial}{\partial p^\nu} \right]S_F(p) = 1 \,.
\label{SFeq}
\end{eqnarray}

The exact solution of this equation gives the Schwinger formula.
As already mentioned, for our purpose it is enough to obtain the
solution only to the linear order in $B$, and therefore we write
it as
\begin{eqnarray}
S_F(p) = S_0(p) + S_B(p) \,,
\label{0+B}
\end{eqnarray}
where $S_0(p)$ is the propagator in the vacuum,
\begin{eqnarray}
S_0(p) = {1 \over \slash p - m + i\epsilon} \,,
\label{S0}
\end{eqnarray}
and $S_B(p)$ is the correction due to the $B$ field.
Substituting this form in \Eq{SFeq} and solving perturbatively,
we then obtain
\begin{eqnarray}
S_B(p) = S_0(p) \left[ \frac{ieQ}{2}F^{\mu\nu}
\gamma_\mu\frac{\partial}{\partial p^\nu} \right] S_0(p) \,.
\label{SB}
\end{eqnarray}
%

\subsection{Propagator of scalars and charged gauge bosons in a
magnetic field} 
Following the approach outlined above, we now find the analogous
expressions for the propagators of charged gauge bosons and
scalars, up to the linear order in $B$.

The scalar field propagator satisfies the equation
\begin{eqnarray}
\left[ {\cal D^\mu} {\cal D_\mu} + m^2 \right]
  \Delta(x,x') = \null - \delta^4(x-x') \,.
\end{eqnarray}
Taking the ansatz
\begin{eqnarray}
\label{scalaransatz}
\Delta(x,x') = \Phi(x,x')\int\,\frac{d^4 p}{(2\pi)^4}
e^{-ip\cdot(x - x')} \Delta_F(p) \,,
\end{eqnarray}
the equation for $\Delta_F(p)$ is 
\begin{eqnarray}
\Big[ \widetilde{\cal D}^\mu \widetilde{\cal D}_\mu -
m^2 \Big]\Delta_F(p) = 1 \,.
\end{eqnarray}
Retaining only up to linear terms in $B$, the equation becomes
\begin{eqnarray}
\label{scalareq}
\left[ p^2 - m^2 - ieQ F_{\mu\nu} p^\mu {\partial \over \partial
    p_\nu} \right] \Delta_F(p) = 1 \,,
\end{eqnarray}
which we solve by writing
\begin{eqnarray}
\Delta(p) = \Delta_0(p) + \Delta_B(p) \,,
\end{eqnarray}
with
\begin{eqnarray}
\Delta_0(p) = {1 \over p^2 - m^2 + i\epsilon} \,.
\label{Delta0}
\end{eqnarray}
Substituting this form in \Eq{scalareq} we then obtain
for the $B$-dependent term
\begin{eqnarray}
\Delta_B(p) = ieQ F_{\mu\nu} p^\mu \Delta_0(p) {\partial \over \partial
    p_\nu} \Delta_0(p) \,.
\label{DeltaB}
\end{eqnarray}

We consider now the charged gauge bosons. The $W$-bosons do not have
minimal couplings with the photon.  In other words, in addition to the
couplings obtained by replacing all partial derivatives in their free
Lagrangian by the EGC derivative defined in \Eq{Dmu}, they have an
anomalous coupling of the form $ W^\dagger_\mu F^{\mu\nu} W_\nu$.
Thus, the terms in the pure gauge Lagrangian involving the quadratic
terms in $W$ and their couplings to the photon can be written in this
suggestive form used by Erdas and Feldman \cite{Erdas:gy}:
\begin{eqnarray}
\lag_{WA} = - \Big({\cal D}_\mu W_\nu \Big)^\dagger 
\Big({\cal D}_\mu W_\nu - {\cal D}_\nu W_\mu \Big) + ie
W^\dagger_\mu W_\nu F^{\mu\nu} \,, 
\label{LWA}
\end{eqnarray}
where $W_\mu$ is the field operator which annihilates the $W^+$
boson.  In addition, we need to introduce the gauge fixing terms which
are necessary for quantizing the gauge fields.  These terms have the
generic form
\begin{eqnarray}
\lag_{\rm gf} = - {1\over \xi} |f_W|^2 \,,
\label{Lfix}
\end{eqnarray}
where $f_W$ contains the $W$-field as well as the unphysical Higgs
boson fields.  In the commonly used $R_\xi$ gauges, one takes
\begin{eqnarray}
f_W = \partial^\mu W_\mu + i\xi M_W \phi^+ 
\label{Rxi}
\end{eqnarray}
so that the gauge fixing Lagrangian contains no interaction term.

In the presence of a background electromagnetic field, however, the pure
gauge Lagrangian contains EGC derivatives of $W$, not simple
derivatives.  Erdas and Feldman \cite{Erdas:gy} pointed out that it is
therefore more convenient that we use a gauge condition that involved
${\cal D}_\mu$ and not just $\partial_\mu$ acting on the $W$-boson
field.  Indeed, such a gauge condition was discussed in a different
context much earlier \cite{Fujikawa:1973qs} where, 
instead of \Eq{Rxi}, the following choice was made,
\begin{eqnarray}
f_W = {\cal D}^\mu W_\mu + i\xi M_W \phi^+ \,.
\label{nonlinf}
\end{eqnarray}
The resulting gauge fixing term, defined
according to \Eq{Lfix}, must be added to the gauge-invariant part of
the Lagrangian before obtaining Feynman rules. An important
consequence of \Eq{nonlinf} is that the resulting Lagrangian has no
cubic coupling involving the $W$-boson, the unphysical Higgs and the
photon.  On the other hand, the Feynman rule for the cubic $WW$-photon
coupling contains the gauge parameter $\xi$, and is given by
\begin{eqnarray}
\begin{picture}(180,70)(-60,-30)
\Photon(-30,30)(0,0){2}{5}
\Text(-17,15)[r]{\rotatebox{-45}{$W^+_\mu(l-k)$}}
\Photon(-30,-30)(0,0){2}{5}
\Text(-17,-15)[r]{\rotatebox{45}{$A_\alpha(k)$}}
\Photon(0,0)(50,0){2}{5}
\Text(25,5)[b]{$W^+_\nu(l)$}
\Text(70,0)[l]{= \quad $ieO_{\alpha\mu\nu}(k,l-k)$}
\SetWidth{2}
\ArrowLine(-16,16)(-14,14)
\ArrowLine(-16,-16)(-14,-14)
\ArrowLine(24,0)(26,0)
\end{picture}
\end{eqnarray}
with
\begin{eqnarray}
O_{\alpha\mu\nu}(k,l-k) = \eta_{\mu\nu} (2l-k)_\alpha  -
\eta_{\nu\alpha} \Big( (2-\zeta) k + \zeta l \Big)_\mu +
\eta_{\alpha\mu} (2k - \zeta l)_\nu  \,,
\label{cubic}
\end{eqnarray}
where $\eta_{\mu\nu}$ is the metric tensor and the shorthand
\begin{eqnarray}
\zeta = 1 - {1 \over \xi} 
\label{zeta}
\end{eqnarray}
has been used.

The equation of motion of the $W$-bosons in the background field, which
is derived from the Lagrangian that follows from
\Eqs{LWA}{Lfix}, together the gauge fixing function defined by
\Eq{nonlinf}, is given by
\begin{eqnarray}
\left[ - \eta_{\alpha\beta} ({\cal D}^2 + M_W^2) + {\cal D}_\alpha
  {\cal D}_\beta - {1 \over \xi} {\cal D}_\beta {\cal D}_\alpha - ie
  F_{\beta\alpha} \right] W^\alpha = 0 \,,
\end{eqnarray}
where the last term comes from the anomalous electromagnetic coupling
of the $W$-bosons that appears in \Eq{LWA}.  Using the
commutation relation
\begin{eqnarray}
\left[ {\cal D}_\alpha, {\cal D}_\beta \right] W^\gamma = ie
F_{\alpha\beta} W^\gamma \,,
\end{eqnarray}
the equation can be rewritten in the form
\begin{eqnarray}
\left[ - \eta_{\alpha\beta} ({\cal D}^2 + M_W^2) + \zeta {\cal
  D}_\beta {\cal D}_\alpha - 2ie F_{\beta\alpha} \right] W^\alpha = 0
  \,,
\end{eqnarray}
and the propagator in the co-ordinate space will then satisfies
the equation
\begin{eqnarray}
\left[ \eta_{\lambda\mu} \left( {\cal D^\alpha}
    {\cal D_\alpha} + M_W^2 \right)- \zeta 
    {\cal D^\lambda} {\cal D^\mu} + 2ieF_{\lambda\mu} \right] 
    D^{\mu\nu} (x,x') = \delta_\lambda^\nu \delta^4(x-x') \,.
\end{eqnarray}
Following the same procedure used above for the fermion and scalar fields,
we obtain the equation for the momentum-space propagator,
\begin{eqnarray}
\left[ \eta_{\lambda\mu} \left( - \widetilde{\cal D}_\alpha
    \widetilde{\cal D}^\alpha + M_W^2 \right) + \zeta 
    \widetilde{\cal D}_\lambda \widetilde{\cal D}_\mu +
    2ieF_{\lambda\mu} \right]  
    D_F^{\mu\nu} (p) = \delta_\lambda^\nu \,,
\end{eqnarray}
where $\widetilde{\cal D}$ has been defined in \Eq{Dp}.
Linearization of this equation gives
\begin{eqnarray}
\left[ 
\eta_{\lambda\mu} (-p^2+M_W^2) 
+ \zeta p_\lambda p_\mu 
- {ie \over 2} F^{\alpha\beta} R_{\alpha\beta\lambda\mu}
\right] D_F^{\mu\nu} (p) = \delta_\lambda^\nu \,, 
\label{DFeq}
\end{eqnarray}
where $R_{\alpha\beta\lambda\mu}$ is defined by
\begin{eqnarray}
R_{\alpha\beta\lambda\mu} = \Big[ -2\eta_{\lambda\mu} p_\alpha + \zeta
  p_\lambda \eta_{\alpha\mu} + \zeta p_\mu \eta_{\alpha\lambda} \Big] 
{\partial \over
\partial p^\beta}   + (\zeta-4) \eta_{\lambda\alpha}
\eta_{\mu\beta} \,,
\end{eqnarray}
and it be expressed in the form
\begin{eqnarray}
R_{\alpha\beta\lambda\mu} = - O_{\alpha\lambda\mu} (0,p) {\partial \over
  \partial p^\beta} 
+ (\zeta-4) \eta_{\lambda\alpha} \eta_{\mu\beta} \,,
\label{R}
\end{eqnarray}
with $O_{\alpha\lambda\mu}$ being the tensor defined in \Eq{cubic}.
As before, we solve \Eq{DFeq} by decomposing the propagator in the form
\begin{eqnarray}
\label{DFmunu}
D_F^{\mu\nu} (p) = D_0^{\mu\nu} (p) + D_B^{\mu\nu} (p)\,,
\end{eqnarray}
where
\begin{eqnarray}
\label{D0}
D^{\alpha\beta}_0(p) = {1 \over p^2 - M_W^2 + i\epsilon} \left( -
\eta^{\alpha\beta} + {(1-\xi) p^\alpha p^\beta \over p^2 - \xi M_W^2}
\right)\,.
\end{eqnarray}
Then, substituting \Eq{DFmunu} into \Eq{DFeq} and
solving for the linear term in $B$ we obtain
\begin{eqnarray}
 D_B^{\mu\nu} (p) = {ie \over 2} F^{\alpha\beta} D_0^{\lambda\mu} (p)
R_{\alpha\beta\lambda\rho} D_0^{\rho\nu} (p) \,.
\end{eqnarray}

In the gauge introduced in \Eq{nonlinf}, the propagator of the
unphysical charged Higgs field in the background magnetic field,
which we denote by $\Delta^{(W)}_F(p)$, can obtained
by making the substitution
\begin{eqnarray}
m^2 \longrightarrow \xi M_W^2
\end{eqnarray}
in the formulas given in \Eqs{Delta0}{DeltaB} for the scalar propagator.
For later reference, we quote the result,
\begin{eqnarray}
\label{DeltaHiggs}
\Delta^{(W)}_F (p) = \Delta^{(W)}_0 (p) + \Delta^{(W)}_B (p)\,,
\end{eqnarray}
where
\begin{eqnarray}
\label{DeltaHiggs0B}
\Delta^{(W)}_0(p) &=& {1 \over p^2 - \xi M_W^2} \,,\nonumber\\
\Delta^{(W)}_B(p) & = & ieQ F_{\mu\nu} p^\mu \Delta^{(W)}_0(p)
{\partial \over \partial p_\nu} \Delta^{(W)}_0(p) \,.
\end{eqnarray}
The Fadeev-Popov ghost propagator should also be modified in a
magnetic field, but we will not need it in our subsequent discussion.

\section{The perturbative approach}\label{s:ptb}
Here we consider the calculation of the $B$-dependent contribution to
the off-shell amplitude for the process $i\stackrel{B}{\rightarrow}f$
using the P method, which we denote by
$\ampl^{(P)}_{i\to f}$.  For clarity, we consider first the case
in which $i$ and $f$ are single particle states (e.g., a neutrino) and
extend the result afterwards to the general case.

\subsection{One-particle amplitude}
\label{subsec:onepartamp}
In the P method the amplitude is expressed in terms of the
off-shell electromagnetic vertex function $\Gamma_\mu(p_i,p_f)$,
which is defined such that the on-shell matrix element
of the electromagnetic current operator $j_\mu(x)$ is given by
\begin{eqnarray}
\label{defgamma}
\langle f(p_f)|j_\mu(0)|i(p_i)\rangle = \bar w_f
\Gamma_\mu(p_i,p_f) w_i \,,
\end{eqnarray}
where $w_{i,f}$ denote the momentum space wavefunctions appropriate
for the particle (e.g., Dirac spinors for fermions, polarization
vectors for spin-1 particles).  The off-shell amplitude for the
transition $i\to f$ in an external electromagnetic field is then given by
\begin{eqnarray}
\ampl^{(P)}_{i \to f}(p_i,p_f) = -i 
\int d^4x \; e^{i(p_f - p_i)\cdot x}A^\mu(x) \Gamma_\mu(p_i,p_f)  \,,
\end{eqnarray}
which can be written as
\begin{eqnarray}
\label{MPonedef}
\ampl^{(P)}_{i \to f}(p_i,p_f) = -i 
\left[\int d^4x \; e^{ik\cdot x}A^\mu(x) \Gamma_\mu(p_i,p_i + k)
\right]_{k = p_f - p_i}  \,.
\end{eqnarray}
For $A(x)$ we now substitute the vector potential given in \Eq{Amu}.
Then using the representation
\begin{eqnarray}
\delta'(t) = {i \over 2\pi} \int_{-\infty}^{+\infty} dz\; ze^{izt}
\end{eqnarray}
as well as the relation
\begin{eqnarray}
\delta' (t - a)f(t) = \null - \delta(t - a) f' (a)
\end{eqnarray}
for the derivative of the delta function, the amplitude is given by
\begin{eqnarray}
\ampl^{(P)}_{i \to f}(p_i,p_f) 
&=& - {1\over 2} (2\pi)^4 \delta^4 (p_f - p_i) F^{\mu\nu} 
\left[{\partial \over \partial k^\nu} \Gamma_\mu(p_i,p_i+k)
\right]_{k = 0} \,,
\label{ampwithgamma}
\end{eqnarray}
where in the last factor we have set $k = 0$
by virtue of the delta function. The $B$-dependent contribution
to the self-energy, which is identified by writing
\begin{eqnarray}
\ampl^{(P)}_{i\to f}(p_i,p_f) = -i (2\pi)^4 \delta^4(p_i-p_f)
\Sigma^{(P)}(p_i) \,, 
\end{eqnarray}
is therefore given by
\begin{eqnarray}
\Sigma^{(P)}(p) = \null - {i \over 2} F^{\mu\nu} 
\left[ {\partial \over \partial k^\nu} \Gamma_\mu(p,p+k)\right]_{k=0} \,.
\label{derivreln}
\end{eqnarray}
%

\subsection{General amplitude}
\label{subsec:generalamp}
We now consider the general case.  Suppose the state $\ket i$
contains $n$ particles with momenta $p_{1,2,...,n}$, and the state
$\ket f$ contains $n'$ particles with momenta $p'_{1,2,...,n'}$. We
denote the total momenta of each state by
\begin{eqnarray}
P & = & \sum_{i = 1}^{n} p_i \,,\nonumber\\
P' & = & \sum_{f = 1}^{n'} p'_f \,.
\end{eqnarray}
In analogy with the one-particle case discussed already, we now define
the off-shell vertex function involving the photon in such a way that
the on-shell matrix element of the electromagnetic current operator
$j_\mu(x)$ between the states $|i\rangle$ and $|f\rangle$ is given by 
\begin{eqnarray}
\langle f|j_\mu(0)|i\rangle = \Big( \bar w_f \Big)^{\alpha'_1\cdots
  \alpha'_{n'}} 
\Gamma_{\mu\alpha_1\cdots\alpha_n \alpha'_1\cdots \alpha'_{n'}}
(p_1,p_2,...,p_n,p'_1,p'_2,...,p'_{n'})  \Big( w_i \Big)^{\alpha_1\cdots
  \alpha_n} \,,  
\end{eqnarray}
where the middle factor is the vertex function, and the other two
factors symbolically denote the collection of momentum space
wavefunctions from the final and initial state particles.  We have put
a general kind of index for these external particles.  For a scalar in
the external states, the corresponding $w$-factor will be unity, and
the index should be absent in the vertex function.  For a fermion
field, we usually suppress the index in favor of a matrix notation.
For vector and tensor fields, the indices appear explicitly.  For the
moment, we will omit all indices on the vertex function except the
photon index, and use the compact notation for the momenta to write
just $\Gamma_\mu(p_i,p'_f)$ for the vertex function, but it should be
understood to be the full quantity defined above, with all indices and
all momenta.

\begin{figure}
\begin{center}
\begin{picture}(70,60)(-35,-40)
\ArrowLine(0,0)(-16,-12)
\ArrowLine(0,0)(-12,-16)
\Text(0,-20)[]{\bf \ldots}
\Text(0,-50)[bc]{(a)}
\Text(20,-20)[]{$P$}
\Text(-20,-20)[]{$P^\prime$}
\ArrowLine(16,-12)(0,0)
\ArrowLine(12,-16)(0,0)
\ArrowArcn(0,10)(10,90,-90)
\Text(12,10)[l]{$l$}
\ArrowArcn(0,10)(10,260,90)
\Text(-14,10)[r]{$l + P - P^\prime$}
\Photon(0,20)(0,40){2}{4}
\GCirc(0,0){5}{.3}
\end{picture}
\hspace{64pt}
\begin{picture}(70,60)(-35,-40)
\ArrowLine(0,0)(-16,-12)
\ArrowLine(0,0)(-12,-16)
\Text(0,-20)[]{\bf \ldots}
\Text(0,-50)[bc]{(b)}
\Text(20,-20)[]{$\tilde P$}
\Text(-20,-20)[]{$\tilde P^\prime$}
\ArrowLine(16,-12)(0,0)
\ArrowLine(12,-16)(0,0)
\ArrowArcn(0,10)(10,90,-90)
\Text(12,10)[l]{$l + q$}
\ArrowArcn(0,10)(10,260,90)
\Text(-14,10)[r]{$l + P - P^\prime$}
\Photon(0,20)(-20,40){1}{5.5}
\Line(0,20)(15,40)
\Line(0,20)(15,35)
\Line(0,20)(15,30)
\Text(20,35)[l]{$\Big\} \Longleftarrow q$}
\GCirc(0,0){5}{.3}
\end{picture}
\end{center}
\caption{Schematic examples of (a) a type-1 diagram, and (b) a type-2
diagram, for the vertex function $\Gamma_\mu(p_i,p_f)$.
The momentum variables are defined in the text.
\label{f:example}
}
\end{figure}
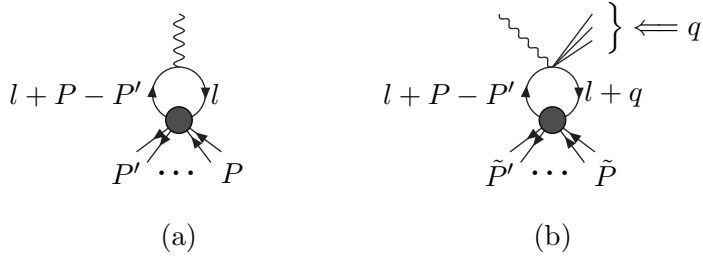

We classify the diagrams that contribute to the vertex function into
two types.  As stated in the Introduction, we deal with processes
where all particles in external states are electrically neutral.
Thus, in the diagrams, the electromagnetic current operator $j_\mu(0)$
is necessarily attached either to an internal line, representing an
electrically charged particle, or to a vertex.  We refer to these
types of diagram as \emph{type-1} and \emph{type-2}, respectively.
Schematic examples of both kinds of diagrams have been shown in
\fig{f:example}.

Let us consider type-1 diagrams first, and denote their contribution
to the vertex function as $\Gamma^{(1)}_\mu$.  In
each such diagram, we can always label the loop-integration momentum
$l$ in such a way that $l$ is the momentum carried by the charged
particle line outgoing from the electromagnetic vertex.  A little
thought reveals that, as a consequence, the line coming in to the same
vertex carries the momentum $l + P - P^\prime$.  The convention is
represented in \fig{f:example}a.  Similarly, we consider a generic
type-2 as shown in \fig{f:example}b.  Here, some of the external
lines are attached to the photon vertex, and we denote by $q$
the net momentum flowing into the diagram due to all these other external
lines. In addition, we denote by $\tilde P$ and $\tilde P^\prime$ the
partial total momentum of the remaining incoming and outgoing lines,
so that $\tilde P - \tilde P^\prime + q = P - P^\prime$.
Then, choosing the integration variable $l$ such that the
momentum carried by the internal line going into the
vertex is again $l + P - P^\prime$, it follows that
the outgoing internal line has momentum $l + q$,
as indicated in \fig{f:example}b.

We now imagine constructing the amplitudes corresponding to each type
of diagram.  For each such amplitude, we define an auxiliary
function by making the replacement
\begin{eqnarray}
\label{introducek}
l + P - P^\prime \rightarrow l - k\,,
\end{eqnarray}
in the propagator (and the tree-level vertex function) of the internal
charged line that goes into the electromagnetic vertex, where $k$ is
an arbitrary vector.  We denote by $\bar\Gamma^{(1)}_\mu(p_i,p_f,k)$
the sum of the auxiliary functions corresponding to the type-1 diagrams,
and similarly by $\bar\Gamma^{(2)}_\mu(p_i,p_f,k)$ the corresponding
sum obtained in the same way for the type-2 diagrams. Furthermore, we define
the total auxiliary vertex function as the sum
\begin{eqnarray}
\bar\Gamma_\mu(p_i,p_f,k) = \bar\Gamma^{(1)}_\mu(p_i,p_f,k) +
\bar\Gamma^{(2)}_\mu(p_i,p_f,k)\,.
\end{eqnarray}
The function $\bar\Gamma_\mu(p_i,p_f,k)$ does not have a direct physical
meaning, but by construction it is such that
\begin{eqnarray}
\bar\Gamma_\mu(p_i,p_f,k)\bigg|_{k = P' - P} =
\Gamma_\mu(p_i,p_f) \,,
\label{Gammakrel}
\end{eqnarray}
which is the important relation for us in what follows.

We want to consider the process of \Eq{i->f}. The off-shell amplitude
in presence of the external magnetic field, which is given by
\begin{eqnarray}
\ampl^{(P)}_{i \to f}(p_i,p_f) = -i 
\int d^4x \; e^{i(P^\prime - P)\cdot x}A^\mu(x)\Gamma_\mu(p_i,p_f)\,,
\end{eqnarray}
can be expressed in terms of the auxiliary vertex function
$\bar\Gamma_\mu(p_i,p_f,k)$ as
\begin{eqnarray}
\ampl^{(P)}_{i \to f}(p_i,p_f) = -i 
\left[\int d^4x \; e^{ik\cdot x}A^\mu(x)\bar\Gamma_\mu(p_i,p_f,k)
\right]_{k = P^\prime - P}\,.
\label{amplB}
\end{eqnarray}
The same manipulations that we applied to \Eq{MPonedef} then
leads us here to an equation that resembles \Eq{ampwithgamma},
\begin{eqnarray}
\ampl^{(P)}_{i \to f}(p_i,p_f) 
= - {1\over 2} (2\pi)^4 \delta^4 (P^\prime - P) F^{\mu\nu}\left\{
{\partial \over \partial k^\nu} \bar\Gamma^{(1)}_\mu(p_i,p_f,k) +
{\partial \over \partial k^\nu} \bar\Gamma^{(2)}_\mu(p_i,p_f,k)
\right\}_{k = 0} \,.
\label{ptbresult}
\end{eqnarray}
%
%
%
It should be remembered that we have been omitting all but the
photon index in writing the $\bar\Gamma^{(1,2)}_\mu$, but in fact they are
assumed to be present on both sides of this equation.
%
%

In spite of the similarity between the type-1 and type-2 contributions
to this equation, there is an important difference between the two types
which a closer look at \fig{f:example} reveals clearly.
If we denote generically by $\tilde S(p)$ the propagator of the
internal line where the electromagnetic vertex is attached (whether
it is a scalar, fermion of vector particle), then in type-1
diagrams that particular propagator appears in a combination
that schematically looks like
\begin{eqnarray}
\label{type1}
\left[\frac{\partial}{\partial k}
\tilde S(l) V\tilde S(l - k)\right]_{k = 0}\,,
\end{eqnarray}
where $V$ is a vertex factor. On the other hand,
for type-2 diagrams the corresponding combination is
\begin{eqnarray}
\label{type2}
\left[\frac{\partial}{\partial k}
\tilde S(l + q)V^\prime \tilde S(l - k)\right]_{k = 0}\,.
\end{eqnarray}

As we will show in the next section, in all cases, whether scalar,
fermion or vector particles, the combination given
in \Eq{type1} can be expressed in terms of the Schwinger
propagator for a particle with momentum $l$,
and that will allow us to prove the equivalence between
the P-method calculation of the type-1 diagrams and the
S-method. On the other hand, no such relation exists
for the combination given in \Eq{type2}, which among
other things depends not just on $l$ but also on $q$.
Therefore, in process for which there are no type-2 diagrams,
the P and the S methods are equivalent.
But otherwise, the S-method does not yield the total
amplitude. The type-2 diagrams of the P-method must
be calculated separately to yield the total amplitude.

\section{Example of equivalence of the two approaches}\label{s:equiv} 
We consider the P-method calculation of the
neutrino self-energy in a background magnetic field.
For the sake of simplicity, we will consider neutrino
interactions with electrons only, which in the 4-Fermi approximation
is given by
\begin{eqnarray}
\mathscr L = \sqrt{2} G_F [\bar e\gamma_\mu(X + Y\gamma_5)e]
[\bar\nu \gamma^\mu L \nu] \,.
\end{eqnarray}
Here $L$ is the projection operator for the left-chiral components of
fermion fields, while $X$ and $Y$ stand for the weak coupling
constants of the electron.
\begin{figure}
\begin{center}
\begin{picture}(120,100)(-60,-50)
\ArrowLine(50,-20)(0,-20)
\ArrowLine(0,-20)(-50,-20)
\Photon(0,20)(0,60){2}{5}
\ArrowArc(0,0)(20,-90,90)
\ArrowArc(0,0)(20,90,270)
\Text(55,-20)[b]{$\nu(p)$}
\Text(-55,-20)[b]{$\nu(p')$}
\Text(-28,0)[br]{$e$}
\Text(28,0)[bl]{$e$}
\Text(6,40)[l]{$\gamma(k)$}
\SetWidth{2}
\ArrowLine(0,41)(0,39)
\end{picture}
\hspace{4cm}
\begin{picture}(120,100)(-60,-50)
\ArrowLine(50,-20)(0,-20)
\ArrowLine(0,-20)(-50,-20)
\Text(55,-20)[b]{$\nu(p)$}
\Text(-55,-20)[b]{$\nu(p)$}
\Text(-28,0)[b]{$e$}
\SetWidth{2}
\ArrowArc(0,0)(20,-90,270)
\end{picture}
\end{center}
\begin{minipage}[t]{0.47\textwidth}
\caption{Diagram for the off-shell vertex function of the neutrino.
  This is required for computing the background field-dependent
  contributions to neutrino self-energy in the perturbative
  approach.}\label{f:nuPer}
\end{minipage}
\hfill
\begin{minipage}[t]{0.47\textwidth}
\caption{Diagram for computing the neutrino self-energy
in a magnetic field.  The thick line indicates that the Schwinger
propagator has to be used.}\label{f:nuSch}
\end{minipage}
\end{figure}

As explained in Section\ \ref{subsec:onepartamp}, in the P-method
we start from the neutrino electromagnetic vertex function
and then determine the $B$-dependent part of the
self-energy by means of \Eq{derivreln}.
In one-loop, the relevant diagram is depicted in \fig{f:nuPer},
which in straightforward fashion yields
\begin{eqnarray}
i\Gamma_\mu (p,p+k) = - \surd 2 G_F \gamma^\alpha L 
  \int {d^4l \over (2\pi)^4} \Tr
\Big[ i\gamma_\alpha (X + Y\gamma_5) iS_0(l) ie\gamma_\mu iS_0(l-k)
  \Big] \,, 
\end{eqnarray}
where, for the electron, we have used $Q=-1$. The $k$ dependence
of the vertex function comes only from the factor $S_0(l - k)$,
and therefore using the relation
\begin{eqnarray}
{\partial \over \partial k^\nu} S_0(l-k) = - {\partial \over
  \partial l^\nu} S_0(l-k) \,,
\label{lkderiv}
\end{eqnarray}
and taking the limit $k \rightarrow 0$, \Eq{derivreln} yields
\begin{eqnarray}
i\Sigma^{(P)} 
&=& \surd 2 G_F \gamma^\alpha L 
  {i \over 2} F^{\mu\nu} \int {d^4l \over (2\pi)^4} \Tr
\Big[ i\gamma_\alpha (X + Y\gamma_5) iS_0(l) ie\gamma_\mu {\partial
    \over \partial l^\nu} iS_0(l) \Big] \,.
\label{sigmaPtb}
\end{eqnarray}

In the S-method, the self-energy is determined to one-loop
from the diagram shown in \fig{f:nuSch}, using the Schwinger propagator
for the internal electron line. In the linear approximation
that we are considering, we use the linear formula for the
propagator given \Eq{0+B}. Thus, denoting by $\Sigma^{(S)}$
the $B$-dependent part of the self-energy calculated in this way,
we obtain
\begin{eqnarray}
i\Sigma^{(S)} = - \surd 2 G_F \gamma^\alpha L
\int {d^4l \over (2\pi)^4} \Tr \Big[ i\gamma_\alpha(X + Y\gamma_5)
  iS_B(l) \Big] \,,
\label{sigmaSch}
\end{eqnarray}
Remembering \Eq{SB}, it follows that $\Sigma^{(S)}$ is identical
to $\Sigma^{(P)}$ given in \Eq{sigmaPtb}. 

Looking closer at the two methods, we can see why they are equivalent
in this case.  In the P-method, the photon vertex appears
between the two electron propagators in the combination
\begin{eqnarray}
C_\mu (k,l) \equiv iS_0(l) \Big( -ieQ\gamma_\mu \Big) iS_0(l - k) \,,
\label{Cmu}
\end{eqnarray}
which using \Eq{lkderiv}, is seen to satisfy
\begin{eqnarray}
\null - {i\over 2} F^{\mu\nu} 
\left[ {\partial \over \partial k^\nu} C_\mu (k,l) \right]_{k=0} =
iS_B(l) \,.  
\label{crucial}
\end{eqnarray}
This Ward-like identity
is the crucial relation that guarantees the equivalence of the
two approaches.  Diagrammatically, it can be represented in the form
\begin{eqnarray}
\null - {i\over 2} F^{\mu\nu} \left[ {\partial \over \partial k^\nu}
  \left( 
  \begin{picture}(70,40)(-35,0)
\Line(0,0)(-16,-12)
\Line(0,0)(-12,-16)
\Text(0,-20)[]{\bf \ldots}
\Line(0,0)(16,-12)
\Line(0,0)(12,-16)
\ArrowArc(0,10)(10,-90,90)
\Text(12,10)[l]{$l-k$}
\ArrowArc(0,10)(10,90,270)
\Text(-12,10)[r]{$l$}
\Photon(0,20)(0,30)22
\ArrowLine(0,31)(0,30)
\Text(0,36)[b]{$k$}
\GCirc(0,0){10}{.3}
  \end{picture}
\right) \right]_{k=0} 
= 
  \begin{picture}(60,50)(-30,0)
\Line(0,0)(-16,-12)
\Line(0,0)(-12,-16)
\Text(0,-20)[]{\bf \ldots}
\Line(0,0)(16,-12)
\Line(0,0)(12,-16)
\CArc(0,10)(10,0,360)
\CArc(0,10)(12,0,360)
\GCirc(0,0){10}{.3}
  \end{picture}
\label{diageqn}
\end{eqnarray}
where the lines at the bottom are external lines, the double line
represents only the magnetic part of the propagator,
and the blob denotes everything else in the diagram.

Let us now consider a more general amplitude that may involve
diagrams in which
the photon line, in the P-method, attaches to an internal scalar line.
Denoting the charge of the scalar by $eQ$,
the factor analogous to the one quoted in \Eq{Cmu} would be in this case
\begin{eqnarray}
C_\mu^\prime (k,l) \equiv i\Delta_0(l) \Big( -ieQ (2l_\mu-k_\mu) \Big)
i\Delta_0(l - k) \,. 
\end{eqnarray}
Although the electromagnetic coupling of the scalar
is momentum-dependent, it still follows that
\begin{eqnarray}
\null - {i\over 2} F^{\mu\nu} 
\left[ {\partial \over \partial k^\nu} C_\mu^{(S)} (k,l) \right]_{k=0} =
i\Delta_B(l) \,,
\label{crucialS}
\end{eqnarray}
as can be simply verified. Thus, the diagrammatic equation of
\Eq{diageqn} applies to this case as well.  Furthermore, this conclusion is
unchanged if the scalar mode is unphysical.

If the photon is attached to an internal $W$-boson line, the factor
analogous to the one quoted in \Eq{Cmu} is given by
\begin{eqnarray}
{C^{\lambda\rho}}_\mu (k,l) \equiv iD^{\alpha\lambda}_0(l) \Big( ie
O_{\mu\alpha\beta} (k,l-k) \Big) iD^{\beta\rho}_0(l - k) \,,
\end{eqnarray}
where the $O_{\mu\alpha\beta}$ is defined in \Eq{cubic}.
Using the analog of \Eq{lkderiv} for the $W$ propagator, we can write 
\begin{eqnarray}
\lim_{k\to0} {\partial \over \partial k^\nu} {C^{\lambda\rho}}_\mu &=& 
-ieD^{\alpha\lambda}_0(l)\lim_{k\to0}  {\partial \over \partial k^\nu}
\left[ O_{\mu\alpha\beta} (k,l-k) D^{\beta\rho}_0(l - k) 
  \right] \nonumber \\ 
&=& -ieD^{\alpha\lambda}_0(l) \left[ - O_{\mu\alpha\beta} (0,l)
  {\partial \over \partial l^\nu} + \left(
  {\partial \over \partial k^\nu} O_{\mu\alpha\beta} (k,l-k)
  \right)_{k=0}  \right] D^{\beta\rho}_0(l)\,,\nonumber\\ 
\end{eqnarray}
and by direct computation using \Eq{cubic} it follows that
\begin{eqnarray}
{\partial \over \partial k^\nu} O_{\mu\alpha\beta} (k,l-k) =
- \eta_{\mu\nu} \eta_{\alpha\beta}  - (2-\zeta)
\eta_{\nu\alpha} \eta_{\mu\beta} + 2 \eta_{\alpha\mu} \eta_{\nu\beta} \,.
\end{eqnarray}
Therefore, contracting with the antisymmetric tensor $F^{\mu\nu}$, we obtain
\begin{eqnarray}
\null - {i\over 2} F^{\mu\nu} \left[ {\partial \over \partial k^\nu}
  {C^{\lambda\rho}}_\mu \right]_{k=0}  = i D^{\lambda\rho}_B \,,
\label{crucialW}
\end{eqnarray}
which proves the equivalence also when the photon is attached to a
$W$-boson line. 
Notice that we have exhausted all the possible ways in which
the photon, in the P-method, can be attached to an internal line
in a diagram since, as emphasized earlier,
there is no $W\phi$-photon trilinear coupling in
the gauge chosen for the $W$'s.

This completes the proof that there is a one-to-one correspondence
between the diagrams in the S-method, and the type-1 diagrams in the
P-method, in which the photon appears attached to the internal
lines of the diagram. Therefore, for transition amplitudes
for which there are no contributions from type-2 diagrams,
both methods give equivalent results.
However, as we will see next, there are
amplitudes for which the P-method involve the type-2, diagrams which have
no counterpart in the S-method. 

\section{Examples of non-equivalence of the two approaches}\label{s:noneq}

\subsection{Processes involving charged gauge bosons}\label{chgb}
\begin{figure}[t]
\begin{center}
\begin{picture}(100,100)(0,-50)
\Photon(0,0)(30,0)25
\ArrowLine(90,-30)(60,-30)
\ArrowLine(60,30)(90,30)
\SetWidth{1.5}
\ArrowLine(30,0)(60,30)
\ArrowLine(60,-30)(30,0)
\Photon(60,30)(60,-30)2{7.5}
\Text(50,-50)[]{\large (a)}
\end{picture}
\quad \quad
\begin{picture}(100,100)(0,-50)
\Photon(0,0)(30,0)25
\ArrowLine(90,-30)(60,-30)
\ArrowLine(60,30)(90,30)
\SetWidth{1.5}
\ArrowLine(30,0)(60,30)
\ArrowLine(60,-30)(30,0)
\DashLine(60,30)(60,-30)4
\Text(50,-50)[]{\large (a$'$)}
\end{picture}
\\
\begin{picture}(100,100)(0,-50)
\Photon(0,0)(30,0)2{5.5}
\ArrowLine(90,-30)(60,-30)
\ArrowLine(60,30)(90,30)
\SetWidth{1.5}
\Photon(30,0)(60,30)2{5.5}
\Photon(60,-30)(30,0)2{5.5}
\ArrowLine(60,-30)(60,30)
\Text(50,-50)[]{\large (b)}
\end{picture}
\quad \quad
\begin{picture}(100,100)(0,-50)
\Photon(0,0)(30,0)2{5.5}
\ArrowLine(90,-30)(60,-30)
\ArrowLine(60,30)(90,30)
\SetWidth{1.5}
\DashLine(30,0)(60,30)4
\DashLine(60,-30)(30,0)4
\ArrowLine(60,-30)(60,30)
\Text(50,-50)[]{\large (b$'$)}
\end{picture}
\end{center}
\caption{One-loop diagrams for the process $Z\to \nu\bar\nu$.  The
  internal solid, wavy and dashed lines represent the electron, the
  $W$-boson and the unphysical charged Higgs, respectively.  In the
  S-method, we should take Schwinger propagators for the
  thick lines.}\label{f:Zdk}
\end{figure}
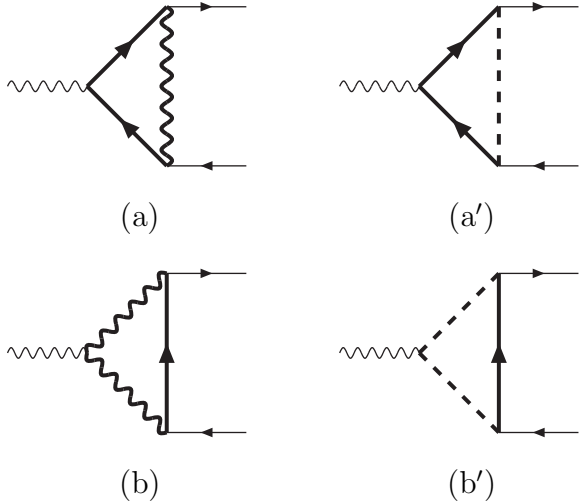

Let us consider the amplitude for the process
\begin{eqnarray}
Z(p) \stackrel{B}{\rightarrow} \nu(p_1) \bar\nu(p_2) \,,
\label{Zdk}
\end{eqnarray}
and focus the attention on the modification to the tree-level,
$B$-independent, term due to the background magnetic field.
Without any of the essential features that are important
for us, we can simplify the discussion by
assuming that the neutrinos are
massless, so that there is no neutrino mixing, and the final state
contains the electron neutrino and its antiparticle.  The one-loop
diagrams which contribute to the amplitude are shown in
\fig{f:Zdk}, where the internal fermion lines represent the electron.

In the S-method, the Schwinger propagators must be used
for the internal lines in these diagrams.
In the linear approximation that we are using,
we need to consider the $B$-dependent part of
the propagator of each line at a time.  

\begin{figure}[t]
\begin{center}
\begin{picture}(100,100)(0,-50)
\Photon(0,0)(30,0)25
\ArrowLine(30,0)(60,30)
\ArrowLine(60,-30)(30,0)
\Photon(60,30)(60,-30)27
\ArrowLine(90,-30)(60,-30)
\ArrowLine(60,30)(90,30)
\Photon(45,15)(15,15)25
\Text(50,-50)[]{\large (a1)}
\end{picture}
\begin{picture}(100,100)(0,-50)
\Photon(0,0)(30,0)25
\ArrowLine(30,0)(60,30)
\ArrowLine(60,-30)(30,0)
\Photon(60,30)(60,-30)27
\ArrowLine(90,-30)(60,-30)
\ArrowLine(60,30)(90,30)
\Photon(45,-15)(15,-15)25
\Text(50,-50)[]{\large (a2)}
\end{picture}
\begin{picture}(100,100)(0,-50)
\Photon(0,0)(30,0)25
\ArrowLine(30,0)(60,30)
\ArrowLine(60,-30)(30,0)
\Photon(60,30)(60,-30)27
\ArrowLine(90,-30)(60,-30)
\ArrowLine(60,30)(90,30)
\Photon(60,0)(90,0)25
\Text(50,-50)[]{\large (a3)}
\end{picture}
\end{center}
\caption{1-loop diagrams for the process $Z + \gamma\to
  \nu\bar\nu$ obtained by attaching a photon line to the diagram of
  \fig{f:Zdk}a.}\label{f:Zdk+ph} 
\end{figure}
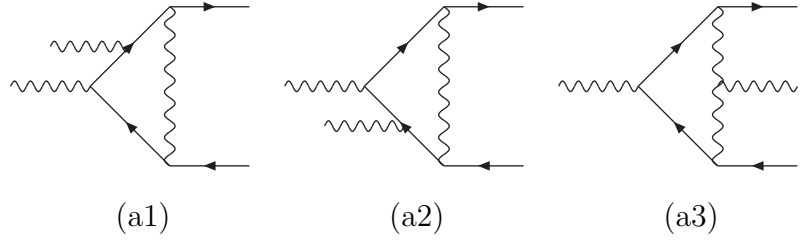
On the other hand, in the P-method, the diagrams are all those
that can be obtained by attaching a
photon to each diagram of Fig.\ \ref{f:Zdk}, in all possible ways.
For example, \fig{f:Zdk+ph} shows all the possible ways
in which a photon line can be
attached to the diagram of \fig{f:Zdk}a.  From the discussion of
Sec.~\ref{s:equiv}, and in particular by means of the identity
in \Eq{crucialW}, it follows that the P-method evaluation
of diagram (a3) of \fig{f:Zdk+ph},
yields a result that is identical to
the contribution that comes from the $B$-dependent part of the 
$W$-boson propagator in the S-method evaluation of \fig{f:Zdk}a.
Similarly, \Eq{crucial} implies that the P-method calculation
of the diagrams (a1) and (a2) of \fig{f:Zdk+ph} is identical
to the contribution that from the $B$-dependent part of the
electron propagator in the S-method evaluation of \fig{f:Zdk}a.
In summary, the S-method evaluation of \fig{f:Zdk}a,
and the P-method evaluation of the 
diagrams (a1), (a2) and (a3) of \fig{f:Zdk+ph}, yield the same result.
The same conclusion holds for \fig{f:Zdk}a$'$ and the corresponding
diagrams of the P-method, since in the latter set of diagrams
the photon is attached to an internal line, 
and the identities in \Eqss{crucial}{crucialS}{crucialW}
can be invoked once again.

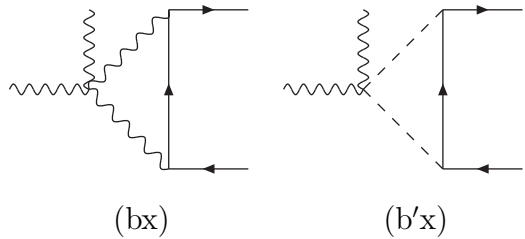
\begin{figure}[b]
\begin{center}
\begin{picture}(100,100)(0,-50)
\Photon(0,0)(30,0)25
\Photon(30,0)(60,30)25
\Photon(60,-30)(30,0)25
\ArrowLine(60,-30)(60,30)
\ArrowLine(90,-30)(60,-30)
\ArrowLine(60,30)(90,30)
\Photon(30,30)(30,0)25
\Text(50,-50)[]{\large (bx)}
\end{picture}
\begin{picture}(100,100)(0,-50)
\Photon(0,0)(30,0)25
\DashLine(30,0)(60,30)4
\DashLine(60,-30)(30,0)4
\ArrowLine(60,-30)(60,30)
\ArrowLine(90,-30)(60,-30)
\ArrowLine(60,30)(90,30)
\Photon(30,30)(30,0)25
\Text(50,-50)[]{\large (b$'$x)}
\end{picture}
\end{center}
\caption{1-loop diagrams for the process $Z + \gamma\to \nu\bar\nu$,
  obtained by attaching a photon line to the $Z$ vertices of the
  diagrams of \fig{f:Zdk}b and \fig{f:Zdk}b$'$.}\label{f:Zdk+phex}
\end{figure}

In the terminology that we have used, the P-method diagrams
that we had to consider so far are type-1 diagrams.
The situation with diagrams \fig{f:Zdk}b and \fig{f:Zdk}b$'$
is different because, in addition to the possibility of
attaching the photon line to each of the internal lines,
that does not exhaust all possibilities.
For example, in
\fig{f:Zdk}b, an extra photon line can be added to the $WWZ$ vertex,
turning it into a $WWZ\gamma$ vertex.  The same can be done to the
diagram in \fig{f:Zdk}b$'$ as well.  
The resulting diagrams, shown in
\fig{f:Zdk+phex}, have no counterpart in the Schwinger method.
In our terminology, they are type-2 diagrams.

We are not showing the type-1 diagrams that correspond to the S-method
evaluation of \fig{f:Zdk}b and \fig{f:Zdk}$'$b,
since they are similar to those
shown in \fig{f:Zdk+ph} and, from the arguments of
Sec.~\ref{s:equiv}, it follows that the evaluation of the
corresponding diagrams in the two methods give the same result once again.

To complete the demonstration of non-equivalence of the two methods,
we now need to show that the contribution of the type-2 diagrams shown
in \fig{f:Zdk+phex} is non-vanishing. Following the
prescription for constructing the auxiliary amplitudes
that correspond to the diagrams, we can write them in the form
\begin{eqnarray}
-i \bar\Gamma_{\mu\nu}^{\rm (bx)} &=& 
 \left({ig \over \sqrt{2}} \right)^2 \int {d^4l \over (2\pi)^4} \Big[
 \gamma_\alpha L iS_0(p-p_2-l) \gamma_\beta L \Big] iD^{\alpha\sigma}_0(l+k)
 iD^{\beta\tau}_0(l-p) iQ_{\sigma\tau\mu\nu} \,, \\ 
-i \bar\Gamma_{\mu\nu}^{\rm (b'x)} &=& 
 \left({igm_e \over \sqrt{2} M_W} \right)^2 \int {d^4l \over (2\pi)^4}
 \Big[ L iS_0(p-p_2-l) R \Big] i\Delta^{(W)}_0(l+k) i\Delta^{(W)}_0(l-p)
 iQ_{\mu\nu} \,,  
\end{eqnarray}
where the propagators of the $W$-boson and the unphysical charge Higgs
boson are given in \Eqs{D0}{DeltaHiggs}. In addition, we have denoted
by $Q_{\sigma\tau\mu\nu}$ and $Q_{\mu\nu}$ the quartic
couplings of the $Z\gamma$ with $WW$ and the unphysical
Higgs which, in the gauge introduced in \Eq{nonlinf},
are given by
\begin{eqnarray}
Q_{\sigma\tau\mu\nu} & = & -eg\cos\theta_W \Big( 2 \eta_{\sigma\tau}
\eta_{\mu\nu} -  \eta_{\sigma\mu} \eta_{\tau\nu} 
- \eta_{\sigma\nu} \eta_{\tau\mu} \Big)\,,\nonumber\\
Q_{\mu\nu} & = & {eg \cos 2\theta_W \over \cos \theta_W} \eta_{\mu\nu}\,,
\end{eqnarray}
respectively. From \Eq{ptbresult}, it then follows that
these diagrams give the following contribution to the amplitude
$Z\to \nu\bar\nu$ amplitude given by
\begin{eqnarray}
\Gamma_\mu^{\rm (x)} &=& - {i\over 2} F^{\nu\lambda} \left[ {\partial
    \over \partial k^\lambda} \Gamma_{\mu\nu}^{\rm (bx+b'x)}
    \right]_{k=0}  \nonumber \\ 
&=& - {g^2\over 4} F^{\nu\lambda} \int {d^4l \over (2\pi)^4} \bigg[ 
 \Big[ \gamma_\alpha L S_0(p_1-l) \gamma_\beta L \Big] 
 {\partial D^{\alpha\sigma}_0(l) \over \partial l^\lambda} 
 D^{\beta\tau}_0(l-p) Q_{\sigma\tau\mu\nu} \nonumber\\*
&& + 
 \left({m_e \over M_W} \right)^2 
 \Big[ L S_0(p_1-l) R \Big] 
 {\partial \Delta^{(W)}_0(l) \over \partial l^\lambda} \Delta^{(W)}_0(l-p)
 Q_{\mu\nu} \bigg] \,,
\end{eqnarray}
where we have made use of relations analogous to \Eq{lkderiv}.
Certainly, this contribution is non-vanishing.  It also shows the
general structure displayed in \Eq{type2}, where the two
propagators appear with different momenta despite the fact that the external
photon momentum has been set to zero.

\newcounter{subd}
\renewcommand{\thesubd}{\alph{subd}}
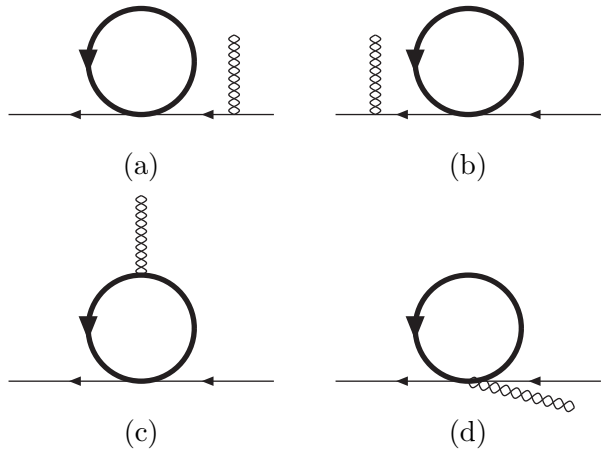
\begin{figure}
\def\eloop{\stepcounter{subd}
\SetWidth{.5}
\ArrowLine(50,0)(0,0)
\ArrowLine(0,0)(-50,0)
\SetWidth{2}
\ArrowArc(0,20)(20,0,360)
\SetWidth{.5}
\Text(0,-20)[]{(\thesubd)}
}
\begin{center}
\begin{picture}(120,100)(-60,-20)
\eloop
\Graviton(35,0)(35,30)25
\end{picture}
%
\begin{picture}(120,100)(-60,-20)
\eloop 
\Graviton(-35,0)(-35,30)25
\end{picture}
\\ 
\begin{picture}(120,100)(-60,-20)
\eloop
\Graviton(0,40)(0,70)25
\end{picture}
%
\begin{picture}(120,100)(-60,-20)
\eloop
\Graviton(0,0)(40,-10)25
\end{picture}
\end{center}
\caption{One-loop diagrams for the process $\nu_1\rightarrow \nu_2 +
  \grav$ in the four-Fermi interaction approximation.
\label{f:nunu'g}
}
\end{figure}
%
\subsection{Theory with linearized gravity}\label{s:lingrav}
In the linear theory of gravity, the couplings of the graviton
with the matter and gauge fields can be determined by 
writing the space-time metric in the form 
\begin{eqnarray}
g_{\lambda\rho} = \eta_{\lambda\rho} + 2\kappa h_{\lambda\rho} \,,
\label{eta+h}
\end{eqnarray}
where $h_{\lambda\rho}$ is identified with the graviton field,
and then expanding the couplings in the Lagrangian
up to the linear order in $\kappa$. The constant $\kappa$
is defined in terms of Newton's gravitational constant by
\begin{eqnarray}
\kappa = \sqrt{8\pi G} \,,
\end{eqnarray}
which is such that the field $h_{\lambda\rho}$ has the properly
normalized kinetic energy term in the Lagrangian.  This point of view
for treating processes involving the gravitational and Standard Model
interactions is the same as that employed in some recent works for the
calculation of quantum gravity amplitudes \cite{Bjerrum-Bohr:2002kt,
Bjerrum-Bohr:2002ks, Nieves:2005ti}, in which General Relativity is
treated as an effective field theory for energies below the Planck
scale \cite{Donoghue:1994dn, Burgess:2003jk}.

The interaction Lagrangian so obtained contains vertices
involving both the photon and graviton that give rise to
type-2 diagrams when we consider processes involving gravitons
in the presence of a $B$ field. In order to make the
discussion concrete, we consider the process
\begin{eqnarray}
\nu_1 \stackrel{B}{\rightarrow} \nu_2 + \grav \,,
\label{nunu'g}
\end{eqnarray}
where $\grav$ is the graviton, and $\nu_1$, $\nu_2$ are two different
neutrino eigenstates.  The couplings involving gravitons that are
relevant to this process have been deduced in the literature
\cite{Choi:1994ax, Shim:1995ap, Nieves:1998xz, Nieves:1999rt,
Nieves:2000dc}. We do not reproduce them here since we will not carry
out the calculation of the amplitude for this process.  We will limit
ourselves to indicate what are the diagrams that are relevant for such
a calculation using the two methods that we have considered.

In the absence of the $B$ field, the amplitude is determined from the
diagrams shown in \fig{f:nunu'g}, the tree-level contribution being
zero due to the fact that the gravitational couplings are flavor
diagonal and universal.  The process cannot occur in the vacuum
because of angular momentum conservation.  But in the background $B$
field, this obstruction is lifted.  We can then try to calculate the
amplitude of the process by interpreting the charged fermion
propagators of \fig{f:nunu'g} as Schwinger propagators.  Further,
although the internal loop can contain all charged leptons, we can
imagine considering only the contributions for the electron in the
loop, with the understanding that restoring the contributions of the
muon and the tau can be made in analogous fashion.  Carrying out the
calculation in this way is the prescription of the S-method.

\begin{figure}
\def\eloop{\stepcounter{subd}
\SetWidth{.5}
\ArrowLine(50,0)(0,0)
\ArrowLine(0,0)(-50,0)
\ArrowArc(0,20)(20,0,360)
}
\begin{center}
\begin{picture}(120,100)(-60,-20)
\eloop
\Photon(0,40)(0,70)25
\Graviton(35,0)(35,30)25
\Text(0,-20)[]{(a)}
\end{picture}
%
\begin{picture}(120,100)(-60,-20)
\eloop 
\Photon(0,40)(0,70)25
\Graviton(-35,0)(-35,30)25
\Text(0,-20)[]{(b)}
\end{picture}
\\ 
\begin{picture}(120,100)(-60,-20)
\eloop
\Photon(-35,50)(-16,32)25
\Graviton(35,50)(16,32)25
\Text(0,-20)[]{(c1)}
\end{picture}
%
\begin{picture}(120,100)(-60,-20)
\eloop
\Photon(35,50)(16,32)25
\Graviton(-35,50)(-16,32)25
\Text(0,-20)[]{(c2)}
\end{picture}
%
\begin{picture}(120,100)(-60,-20)
\eloop
\Photon(0,40)(0,70)25
\Graviton(0,0)(40,-10)25
\Text(0,-20)[]{(d)}
\end{picture}
\\
\begin{picture}(120,100)(-60,-20)
\eloop
\Photon(0,40)(-20,70)25
\Graviton(0,40)(20,70)25
\Text(0,-20)[]{(e)}
\end{picture}
\begin{picture}(120,100)(-60,-20)
\eloop
\Photon(0,40)(0,70)25
\Graviton(0,55)(35,55)25
\Text(0,-20)[]{(f)}
\end{picture}
\end{center}
\caption{One-loop diagrams for the $\nu_1 + \gamma \rightarrow \nu_2 +
  \grav$ process in the 4-Fermi approximation. In our terminology,
the diagrams (a)-(-d) are type-1 diagrams while (e) and (f) are type-2.}
\label{f:nupnu'g}
\end{figure}
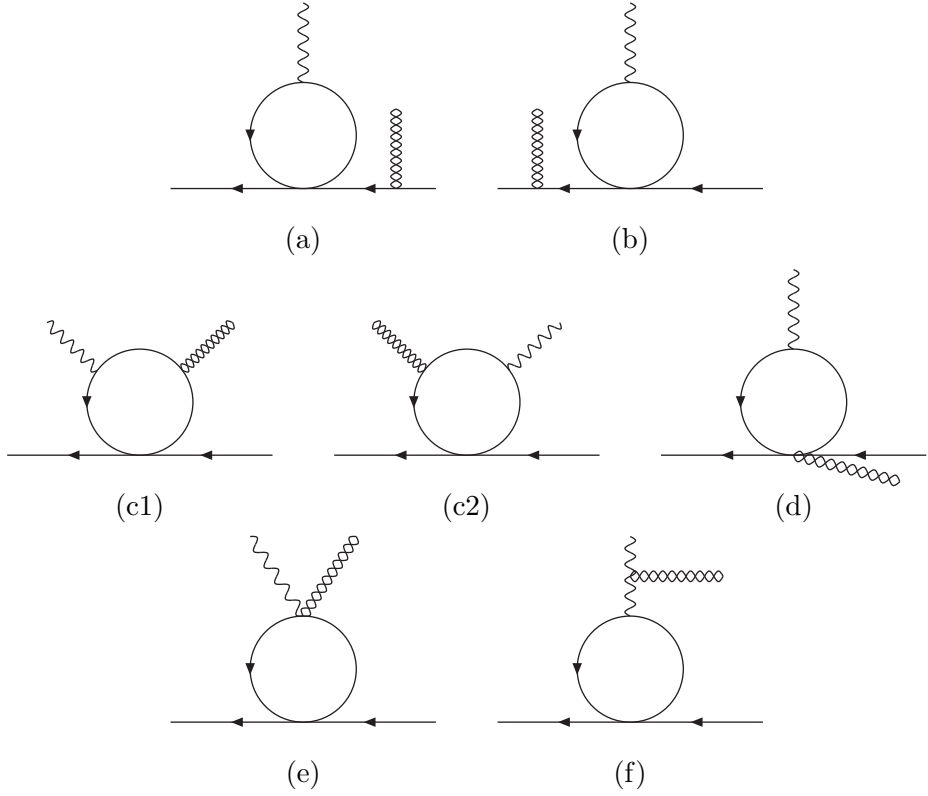
Let us now shift our attention to the P-method and look at the process
\begin{eqnarray}
\nu_1 + \gamma \to \nu_2 + \grav \,,
\label{nupnu'g}
\end{eqnarray}
for which the relevant diagrams are shown in \fig{f:nupnu'g}.
Clearly, diagrams (a)-(d) contain the usual QED vertex for the
electron and therefore the factor $C_\mu$ given in \Eq{Cmu}.  The
calculation of the amplitudes of these diagrams yields the same result
as that obtained from the S-method calculation of the diagrams of
\fig{f:nunu'g}.

However, the contribution of the type-2 diagrams
(e) and (f) of \fig{f:nupnu'g} have no counterpart in the S-method.
If we do not consider their contribution to the total amplitude,
then if we consider the process of \Eq{nupnu'g} as a physical process
involving a real photon, the amplitude does not satisfy
the transversality requirements due to the gravitational and
electromagnetic gauge invariance. 
Moreover, the total amplitude, taking those two diagrams
into account does satisfy the aforementioned conditions.
We have verified this explicitly. 
This reinforces our conclusion that the S-method does not yield
the total amplitude in those cases in which the P-method involves
type-2 diagrams.

\section{Conclusions}\label{s:conclu}
In this work we have considered the calculation of amplitudes
for processes that take place in a constant background magnetic field $B$
with the purpose of comparing two methods.
One, to which refer as the P-method, uses the standard method
for the calculation of an amplitude
in an external field. The other, the S-method, utilizes the
Schwinger propagator for charged particles in a magnetic field.

We showed that there are processes for which the two methods
of calculating the amplitude yield equivalent results. We illustrated
this with the specific example of the neutrino forward scattering
amplitude in a magnetic field, and we indicated specifically the
propagator identities that operate to guarantee the equivalence
in that case.

However, we pointed out that there are processes
for which the Schwinger propagator method does not yield the total
amplitude. For illustrative purposes, we considered the amplitude for $Z$ decay
into a neutrino-antineutrino pair in a $B$ field, in the context 
of the standard model.
In that case, the diagrams that must be included in the P-method of calculation
can be divided in two groups, that we called type-1 and type-2 diagrams.
In the type-1 diagrams the electromagnetic vertex is attached to an internal
propagator line, while in type-2 diagrams there are some external lines as well
attached to that vertex. We showed that there is a one-to-one correspondence
between the diagrams of the S method and the type-1 diagrams of the P-method
and that their calculations yield the same result.
Therefore, for processes for which the type-2 diagrams do not exist,
both the P and S methods yield the same result, which is the case
of the neutrino processes that we have mentioned. 
However, the type-2 diagrams have no counterpart in the S-method and therefore
this method does not yield the complete amplitude. The
total amplitude is obtained by taking the result of the type-1 diagrams,
which can be calculated by either method, and then adding
the result of the type-2 diagrams using the P method.

Moreover, we indicated that leaving out the type-2 diagrams in the
calculation of the amplitude does not yield a gauge-invariant result.
We have verified this explicitly by considering the neutrino
decay into another neutrino and a graviton in a $B$ field, which is another
process for there are type-2 diagrams.
In that particular case, we also verified that by including the type-2
diagrams the gauge invariance of the amplitude is restored,
which reinforces our conclusion that the S-method does not yield
the total amplitude in the general case in which the P-method
contains type-2 diagrams.

Needless to say, in the original context of the Schwinger propagator,
namely QED, there are no type-2 diagrams, so that both methods 
are equivalent. However, the situation is different when the
other particles and interactions of the standard model are taken
into account, and/or possibly the gravitational interactions
as well. Our remarks should be taken within these broader contexts.

\subsection*{Acknowledgements}
The work of JFN was supported by the U.S. National Science
Foundation under Grant 0139538. 


\end{document}